\documentclass[aps,prd,preprint,superscriptaddress,preprintnumbers,nofootinbib,tightenlines,floatfix]{revtex4-1}

\pdfoutput=1 

\usepackage{graphicx}
\usepackage{dcolumn}
\usepackage{bm}
\usepackage[figuresleft]{rotating}
\usepackage{color}

\newcolumntype{d}{D{.}{.}{-1}}
\newcommand{\etaenu}{D^+ \to \eta e^+ \nu_e}

\newcommand{\etapenu}{D^+ \to \eta^\prime e^+ \nu_e}

\newcommand{\phienu}{D^+ \to \phi\, e^+ \nu_e}
\newcommand{\mcc}[1]{\multicolumn{1}{c}{#1}}

\newcommand{\vcd}{|V_{cd}|}

\newcommand{\dele}{\Delta E}
\newcommand{\bcm}{M_{\text{BC}}}

\newcommand{\psip}{\psi(2S)}
\newcommand{\psipp}{\psi(3770)}

\newcommand{\EE}{e^+e^-}

\newcommand{\pip}{\pi^+}
\newcommand{\pim}{\pi^-}
\newcommand{\piz}{\pi^0}
\newcommand{\pipi}{\pip\pim}

\newcommand{\ra}{\rightarrow}

\newcommand{\ddb}{D\bar{D}}

\newcommand{\Dp}{D^+}
\newcommand{\Dm}{D^-}
\newcommand{\enu}{e^+ \nu_e}

\newcommand{\mbc}{M_{\text{BC}}}
\newcommand{\ks}{K_S^0}
\newcommand{\kl}{K_L^0}
\newcommand{\etap}{\eta^\prime}

\newcommand{\GG}{\gamma\gamma}

\newcommand{\KK}{K^+K^-}

\newcommand{\bmath}{\begin{displaymath}}
\newcommand{\emath}{\end{displaymath}}

\newcommand{\beq}{\begin{equation}}
\newcommand{\eeq}{\end{equation}}
\newcommand{\bfg}{\begin{figure}}
\newcommand{\efg}{\end{figure}}
\newcommand{\bitm}{\begin{itemize}}
\newcommand{\eitm}{\end{itemize}}
\newcommand{\bnum}{\begin{enumerate}}
\newcommand{\enum}{\end{enumerate}}
\newcommand{\btbl}{\begin{table}}
\newcommand{\etbl}{\end{table}}
\newcommand{\btbu}{\begin{tabular}}
\newcommand{\etbu}{\end{tabular}}

\newcommand{\emissev}{\ensuremath{E_{\rm miss}^{\rm evt}}}
\newcommand{\pmissev}{\ensuremath{{\bf p}_{\rm miss}^{\rm evt}}}

\newcommand{\ExcludeThisText}[1]{}

\mathchardef\mhyphen="2D

\begin{document}

\preprint{CLNS 10/2067}  
\preprint{CLEO 10-4 }    

\title{\boldmath Studies of $D^+ \rightarrow \{\etap,\eta,\phi\} e^+ \nu_e $}

\author{J.~Yelton}
\affiliation{University of Florida, Gainesville, Florida 32611, USA}
\author{P.~Rubin}
\affiliation{George Mason University, Fairfax, Virginia 22030, USA}
\author{N.~Lowrey}
\author{S.~Mehrabyan}
\author{M.~Selen}
\author{J.~Wiss}
\affiliation{University of Illinois, Urbana-Champaign, Illinois 61801, USA}
\author{M.~Kornicer}
\author{R.~E.~Mitchell}
\author{M.~R.~Shepherd}
\author{C.~M.~Tarbert}
\affiliation{Indiana University, Bloomington, Indiana 47405, USA }
\author{D.~Besson}
\affiliation{University of Kansas, Lawrence, Kansas 66045, USA}
\author{T.~K.~Pedlar}
\author{J.~Xavier}
\affiliation{Luther College, Decorah, Iowa 52101, USA}
\author{D.~Cronin-Hennessy}
\author{J.~Hietala}
\author{P.~Zweber}
\affiliation{University of Minnesota, Minneapolis, Minnesota 55455, USA}
\author{S.~Dobbs}
\author{Z.~Metreveli}
\author{K.~K.~Seth}
\author{A.~Tomaradze}
\author{T.~Xiao}
\affiliation{Northwestern University, Evanston, Illinois 60208, USA}
\author{S.~Brisbane}
\author{J.~Libby}
\author{L.~Martin}
\author{A.~Powell}
\author{P.~Spradlin}
\author{G.~Wilkinson}
\affiliation{University of Oxford, Oxford OX1 3RH, UK}
\author{H.~Mendez}
\affiliation{University of Puerto Rico, Mayaguez, Puerto Rico 00681}
\author{J.~Y.~Ge}
\author{D.~H.~Miller}
\author{I.~P.~J.~Shipsey}
\author{B.~Xin}
\affiliation{Purdue University, West Lafayette, Indiana 47907, USA}
\author{G.~S.~Adams}
\author{D.~Hu}
\author{B.~Moziak}
\author{J.~Napolitano}
\affiliation{Rensselaer Polytechnic Institute, Troy, New York 12180, USA}
\author{K.~M.~Ecklund}
\affiliation{Rice University, Houston, Texas 77005, USA}
\author{J.~Insler}
\author{H.~Muramatsu}
\author{C.~S.~Park}
\author{L.~J.~Pearson}
\author{E.~H.~Thorndike}
\author{F.~Yang}
\affiliation{University of Rochester, Rochester, New York 14627, USA}
\author{S.~Ricciardi}
\affiliation{STFC Rutherford Appleton Laboratory, Chilton, Didcot, Oxfordshire, OX11 0QX, UK}
\author{C.~Thomas}
\affiliation{University of Oxford, Oxford OX1 3RH, UK}
\affiliation{STFC Rutherford Appleton Laboratory, Chilton, Didcot, Oxfordshire, OX11 0QX, UK}
\author{M.~Artuso}
\author{S.~Blusk}
\author{R.~Mountain}
\author{T.~Skwarnicki}
\author{S.~Stone}
\author{J.~C.~Wang}
\author{L.~M.~Zhang}
\affiliation{Syracuse University, Syracuse, New York 13244, USA}
\author{G.~Bonvicini}
\author{D.~Cinabro}
\author{A.~Lincoln}
\author{M.~J.~Smith}
\author{P.~Zhou}
\author{J.~Zhu}
\affiliation{Wayne State University, Detroit, Michigan 48202, USA}
\author{P.~Naik}
\author{J.~Rademacker}
\affiliation{University of Bristol, Bristol BS8 1TL, UK}
\author{D.~M.~Asner}
\altaffiliation[Now at: ]{Pacific Northwest National Laboratory, Richland, WA 99352}
\author{K.~W.~Edwards}
\author{K.~Randrianarivony}
\author{G.~Tatishvili}
\altaffiliation[Now at: ]{Pacific Northwest National Laboratory, Richland, WA 99352}
\affiliation{Carleton University, Ottawa, Ontario, Canada K1S 5B6}
\author{R.~A.~Briere}
\author{H.~Vogel}
\affiliation{Carnegie Mellon University, Pittsburgh, Pennsylvania 15213, USA}
\author{P.~U.~E.~Onyisi}
\author{J.~L.~Rosner}
\affiliation{University of Chicago, Chicago, Illinois 60637, USA}
\author{J.~P.~Alexander}
\author{D.~G.~Cassel}
\author{S.~Das}
\author{R.~Ehrlich}
\author{L.~Fields}
\author{L.~Gibbons}
\author{R.~Gray}
\altaffiliation[Now at: ]{Rutgers University, Piscataway, New Jersey 08855, USA}
\author{S.~W.~Gray}
\author{D.~L.~Hartill}
\author{B.~K.~Heltsley}
\author{D.~L.~Kreinick}
\author{V.~E.~Kuznetsov}
\author{J.~R.~Patterson}
\author{D.~Peterson}
\author{D.~Riley}
\author{A.~Ryd}
\author{A.~J.~Sadoff}
\author{X.~Shi}
\author{W.~M.~Sun}
\affiliation{Cornell University, Ithaca, New York 14853, USA}
\collaboration{CLEO Collaboration}
\noaffiliation


\date{November 4, 2010}

\begin{abstract}

We report the first observation of the decay $\Dp\ra\etap\enu$ in two analyses,
which combined provide a branching fraction of 
${\cal B}(D^{+}\ra\etap\enu) = (2.16\pm0.53\pm0.07)\times 10^{-4}$.  We also provide an improved measurement of
${\cal B}(D^{+}\ra\eta\enu) = (11.4\pm0.9\pm0.4)\times 10^{-4}$, provide
the first form factor measurement, and set the improved upper limit
${\cal B}(\phienu) < 0.9\times 10^{-4}$ (90\%~C.L.).
\end{abstract}

\maketitle

Semileptonic decay provides an excellent laboratory for the study of both weak and strong interactions.
Charm semileptonic decay allows determination of the parameters
$|V_{cd}|$ and $|V_{cs}|$
from the Cabibbo-Kobayashi-Maskawa (CKM) matrix~\cite{ckm},
and stringent testing of predictions for QCD contributions to the decay amplitude.
A complete understanding of charm semileptonic decay requires
study of both high-statistics and rare modes.

The semileptonic decay $\Dp \to \etap\enu$ has not yet been observed. 
Its rate relative to $\Dp \to \eta\enu$ will
provide information about $\eta$-$\eta^\prime$ mixing~\cite{isgw2},
as well as about the role of the QCD anomaly in heavy quark decays
involving $\eta'$~\cite{etap:FKS}. Study of these modes
also probes the composition of the $\eta$
and $\eta^\prime$ wave functions when combined with measurements
of the corresponding $D_s$ semileptonic decay modes~\cite{Bigi}, and can gauge the
possible role of weak annihilation in the corresponding
$D_s$-meson semileptonic decays. The process $\phienu$ 
is not expected to occur in the absence of mixing between
the $\omega$ and $\phi$.

The differential decay rate for $D^{+}\to\eta e^{+}\nu_{e}$ is given, in the limit of negligible electron mass,  by
\beq
\label{eq:integ}
\frac{d\Gamma}{dq^2} = \frac{G_F^2|V_{cd}|^2 |{\bf p}_{\eta}|^3}{24 \pi^3}|f_+(q^2)|^2,
\eeq
where $G_F$ is the Fermi constant, $V_{cd}$ is the CKM matrix element for $c\to d$ quark transitions,
and ${\bf p}_{\eta}$ is the $\eta$ momentum in the $D$ meson's rest frame.  The form factor $f_+(q^2)$
parametrizes the strong interaction dynamics as a function of the hadronic four-momentum
transfer $q^{2}$.
By measuring the partial branching fraction
as a function of $q^{2}$, we probe $f_+(q^2)$, providing a test of the theoretical framework for
calculation of the form factors needed to determine many CKM matrix elements.

We report herein on the first observation  of $\Dp\ra\etap\enu$ and a measurement of its
branching fraction,
on an improved measurement of ${\cal B}(\Dp \to \eta\enu)$ and first measurement
of its form factor,
and on an improved search for $\phienu$. Charge-conjugate modes are implied throughout this article.
The  results derive from two analyses of $818~\text{pb}^{-1}$ of $e^{+}e^{-}$ collision data
collected with the CLEO-c detector~\cite{cleo_detector}
at the $\psipp$ resonance. 
The data include $\sim 2.4\times 10^6$ $\Dp\Dm$ events.

%

One analysis employs the tagging technique used in past CLEO-c studies of
these~\cite{281etaenu} and other~\cite{818kpienu,Asner:2009pu} semileptonic modes.
A parent event sample is defined by reconstruction of either $D^{\pm}$ meson
in a specific hadronic decay mode (the tag).
The fraction of these parent events in which the other $D$ is reconstructed in the signal semileptonic mode determines the absolute semileptonic branching fraction 
${\cal{B}}_{{\rm SL}} =  (N_{ \rm tag, SL }/N_{ \rm tag})(\epsilon_{ \rm tag}/\epsilon_{\rm tag, SL})$. Here $N_{\rm tag}$ and $\epsilon_{\rm tag}$ are the yield and reconstruction efficiency, respectively, for the hadronic tag, 
and $N_{{\rm tag, SL}}$ and  $\epsilon_{\rm tag, SL}$ are those for the combined semileptonic decay and hadronic tag~\cite{281etaenu}.  

The six tag modes
$K^+ \pim\pim$, $K^+ \pim\pim\piz$, $\ks\pim$, $\ks \pim\piz$, $\ks \pim\pim\pip$, and $\KK\pim$ 
are selected based on 
the difference in energy $\dele \equiv E_D - E_{\text{beam}}$ of the
$D$ tag candidate ($E_D$) and the beam
($E_{\text{beam}}$), and on the beam-constrained mass $\mbc \equiv
(E^2_{\text{beam}}/c^4 - |{\bf p}_D|^2/c^2)^{1/2}$, where
${\bf p}_D$ is the reconstructed momentum of the $D$ candidate.
 Reference~\cite{Dobbs:2007zt} summarizes the selection criteria and their performance for the 
 $\pi^{\pm}$, $K^{\pm}$, 
 $\piz$, and $\ks$ candidates.
From multiple candidates of the same mode and charge, we choose
that with the smallest $|\Delta E|$.
The yield of each tag mode is obtained from a fit~\cite{Dobbs:2007zt} to its $\mbc$ distribution.
We find a total of 481223$\pm$809 $D^{\pm}$ tags.

We search each tagged event for an $e^{+}$ and an
$\eta$ ($\gamma\gamma$ and $\pipi\piz$ modes), $\etap$ ($\pipi\eta$ and $\rho\gamma$ modes), 
or $\phi$ ($K^{+}K^{-}$ mode) candidate following Ref.~\cite{281etaenu}. 
Candidate $\pi^{\pm}e^{\mp}$ or $K^{\pm}e^{\mp}$ pairs must have an opening angle $\theta > 20^\circ$ to suppress $\gamma$ conversion
backgrounds.
For $\etap\to\rho\gamma$ candidates,
the $\pi^{\pm}$ and $\gamma$ must have 
 an opening angle $\theta_{\pi\gamma}$ in the $\rho^{0}$ rest frame satisfying
$|\cos\theta_{\pi\gamma}|<0.70$. Signal varies as $\sin^{2}\theta_{\pi\gamma}$,
while background is flat.
The combined tag and semileptonic candidates must
account for all tracks in the event.
The undetected neutrino leads to missing energy
$E_{\text{miss}}\equiv E_{\rm beam}-E_{h+e}$ and missing momentum
${\bf p}_{\text{miss}}\equiv-[{\bf p}_{h+e}+\widehat{\mathbf{p}}_{\rm tag}((E_{\rm beam}/c)^2-m^2_{D}c^{2})^{1/2}]$,
where $E_{h+e}\equiv E_{h}+E_{e}$, ${\bf p}_{h+e}\equiv {\bf p}_{h}+{\bf p}_{e}$, and $\widehat{\mathbf{p}}_{\rm tag}$ is the unit vector in the
direction of the tag $D^{-}$ momentum.
Correctly reconstructed semileptonic candidates
peak at zero in 
$U \equiv E_{\text{miss}} - c|{\bf p}_{\text{miss}}|$,
which has a 10 MeV resolution. 
For each tag mode of a given charge in an event, we allow only one semileptonic candidate. We take the candidate with the smallest $\sum_{X} \chi^{2}_{M}(X)$,
where we sum over all reconstructed $X\in(\piz, \eta, \etap, \phi)$ particles in a candidate.
The pull
$\chi_{M}(X)\equiv (M_{\rm r}-M_{X})/\sigma_{M}$, 
where $M_{\rm r}$ and  $M_{X}$ are the reconstructed and nominal~\cite{pdg2008} masses for particle $X$, and
the resolution $\sigma_{M}$ derives from the error matrices of the daughters of $X$.

\begin{figure*}[tb]
\includegraphics*[width=12cm]{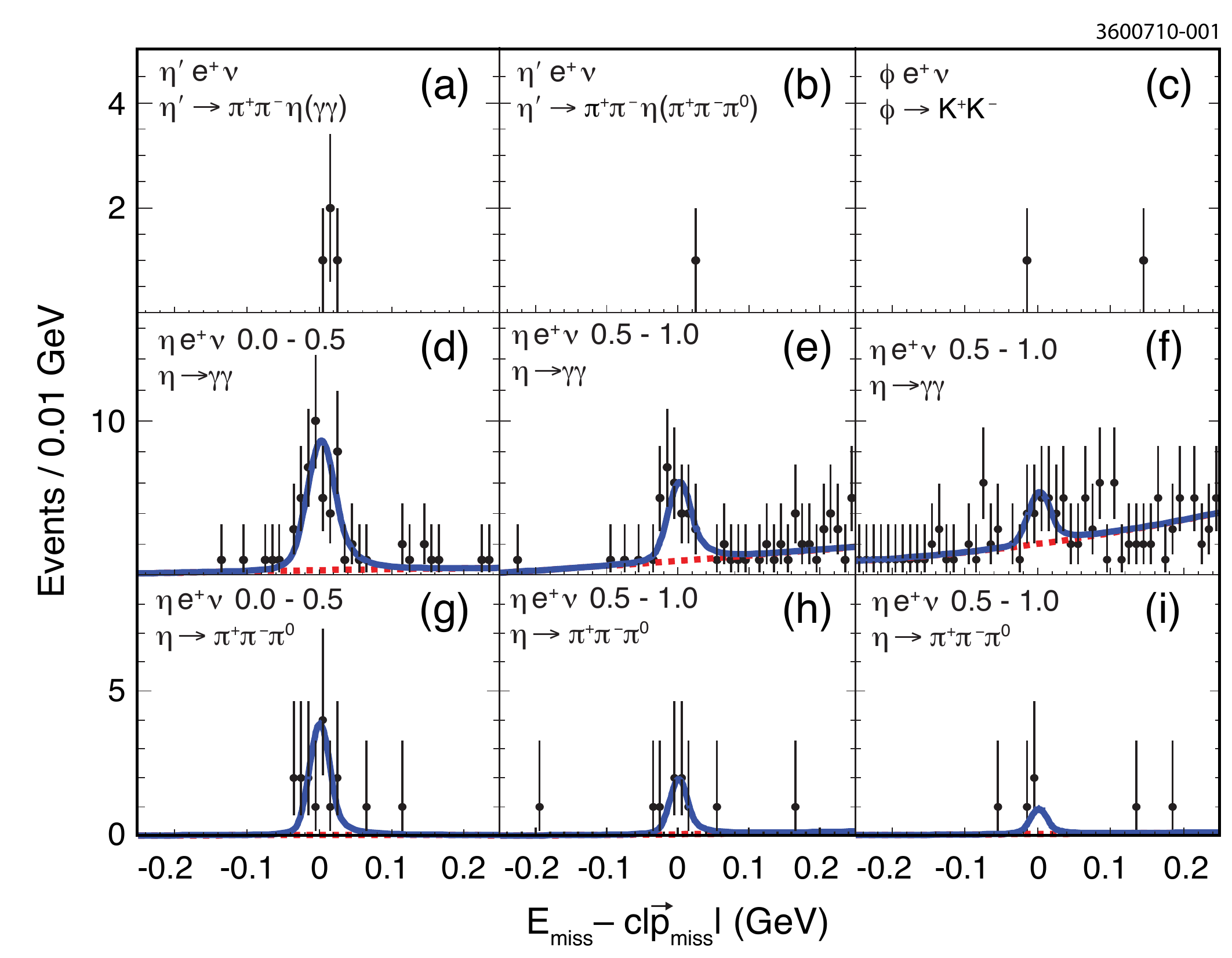}
\caption{\label{fig:U} Tagged analysis $U$ distributions in data (points) for
$\Dp\ra\etap\enu$ (a,b),  $\Dp\ra\phi\enu$ with $\phi\ra\KK$ (c), and
$\Dp\ra\eta\enu$
with $\eta\ra\GG$ (d--f) and $\eta\ra\pipi\piz$ (g--i),
in the three $q^2$ intervals. The total (solid line)  and background (dashed line) 
distributions from the fits are also shown.
}
\end{figure*}


The second analysis, generic reconstruction (GR) \cite{RichardThesis}, 
refines techniques optimized for association of event-wide missing energy (\emissev) and momentum (\pmissev)
with a neutrino~\cite{Nadia}.  We apply the track and photon selection algorithms of Ref.~\cite{Nadia},
and impose associated event-level criteria to reduce
background from undetected particles: 
the charges of the selected tracks must sum to zero and the number of identified $e^{\pm}$ must be exactly one.
We then search for $\eta$ ($\gamma\gamma$, $\pipi\piz$, and $3\piz$ modes), $\etap$ 
($\pipi\eta$, $\piz\piz\eta_{\gamma\gamma}$, $\rho\gamma$, and $\gamma\gamma$ modes) candidates, using criteria~\cite{RichardThesis}
similar to those of the tagged analysis. This analysis requires
$|\cos\theta_{\pi\gamma}|<0.85$. 

For each 
$\eta^{(')}e^{\pm}\nu_{e}$
candidate, the GR algorithm attempts reconstruction of a hadronic decay for the 
second $D$ from the remaining particle content, a departure from the
previous neutrino reconstruction measurements. Doing so both improves the \emissev\ and \pmissev\ resolutions and suppresses combinatoric background.  
The second $D$ reconstruction begins with a closer examination of the remainder of the tracks in the event, which, again,
happens separately for {\em each} semileptonic candidate in an event.

From the selected tracks that are not used in the semileptonic candidate,
we form two sets: 
(i) non-overlapping $\ks\to\pipi$ candidates, and
(ii) tracks consistent with
originating from the the primary $e^{+}e^{-}$ interaction.
The $\ks$ candidates must be within 12 MeV$/c^{2}$ of $M_{K^{0}}$,
overlapping $\ks$ candidates are resolved using the best mass, and final $\ks$ candidates are kinematically fit with a mass constraint.
A track is consistent with the primary interaction vertex if it is consistent with the beam envelope (within 5 cm of the origin
along the beam direction and within 0.5 cm radially).
A selected track outside of these categories is most likely a $\ks$ daughter whose sibling was used in the semileptonic candidate, so that candidate is rejected.

To enhance photon candidate purity, we also form a set of non-overlapping $\pi^{0}\to\gamma\gamma$ and $\eta\to\gamma\gamma$ 
candidates.  Overlaps are resolved based on the smallest magnitude $\chi_{M}(\pi^{0})$ or $\chi_{M}(\eta)$.  
The algorithm's need for high efficiency dictates that we allow the broad ranges $-25 < \chi_{M} < 15$ for $\pi^{0}$  candidates
and $-15 < \chi_{M} < 15$ for $\eta$ candidates.
Unpaired showers with energy below
100 MeV are likely remnants from hadronic shower and are vetoed.  The veto energy is raised to
250 MeV if any $K^{\pm}$ candidate is found in the event.  The  $\gamma\gamma$
candidates are kinematically refit with a mass constraint prior to use in the 
reconstruction of the second $D$ and the neutrino.

The $\pi^{0}$, $\eta$, and $\ks$ candidates, along with the remaining photons and tracks, form the second, non-signal (ns), $D$
candidate with momentum
${\bf p}_{\rm ns}$ and energy $E_{\rm ns}$. They are further combined with
signal $e^{\pm}$ and $\eta^{(\prime)}$ candidates and compared with the total four momentum of the electron-positron
collision to
estimate \emissev\ and \pmissev. The signal $D$ momentum (${\bf p}_{\rm sig}$) and energy ($E_{\rm sig}$) can then be reconstructed.  
The signal and non-signal $D$ candidates must have opposite sign; the signal $e^{\pm}$
and the non-signal $D$ daughters must respect charge correlation assuming Cabibbo-favored decays. We require
$\emissev > 50$ MeV and a total vetoed-shower 
energy under 300 MeV.  
$\Delta E$ for both $D$ candidates and $\emissev-c|\pmissev|$ 
must be consistent with
zero within mode-dependent limits of about 100 MeV.
To improve
resolution in $\Delta E_{\rm sig}$, we take $E_{\nu}=c|\pmissev|$.  
By making the further, very good, assumption that the $|\pmissev|$ resolution 
dominates the $\Delta E_{\rm sig}$ resolution, we can also improve $M_{BC}$.
We rescale \pmissev by a correction $\zeta$ that would result in
$\Delta E_{\rm sig}=0$:  ${\bf p}_{\nu}=\zeta\pmissev$ with $\zeta=1+\Delta E_{\rm sig}/(c|\pmissev|)$.
Signal mode yields are determined from fits to the resulting $\bcm$ distributions.
To increase signal
sensitivity in our yield fit, we classify a high-quality (HQ) sample with the following properties: no unused showers, all 
$\gamma\gamma$ candidates with $-5<\chi_{M}(Y)<3$, $Y\in (\eta, \pi^{0})$, and a non-signal $D$ satisfying the tagged analysis
$\Delta E$ and $\bcm$ criteria.
Reconstruction efficiencies, not including submode branching fractions, range from 2--5\% overall, and 1--3\% for the
HQ subsample.

To reduce the dominant source of background, misreconstructed decays of other more copious charm semileptonic modes, 
the GR $\eta^{\prime}$ candidates must satisfy  $\chi^{2}_{M}(D)_{\eta^{\prime}}-\chi^{2}_{M}(D)_{\rm min}<9$, where $\chi_{M}(D)_{\rm min}$ is the smallest magnitude non-signal $D$ mass pull of all semileptonic candidates in an event.
The additional charged and neutral $D\to X e\nu$ modes considered for the requirement include 
$X=\pi^{\pm}$, $\pi^{0}$, $K^{\pm}$, $K^{*\pm}$, $K^{*0}$, $K_{S}$ ($\pipi$ mode), $\rho^{\pm}$, and $\rho^{0}$.    
This requirement halves the background level with 90\% signal efficiency.  

\begin{figure*}[tb]
\includegraphics*[width=10.5cm]{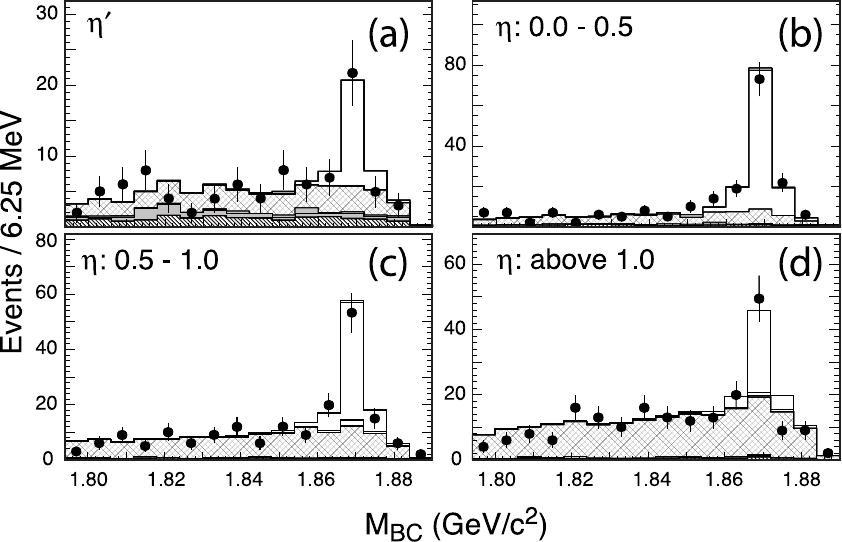}
\caption{\label{fig:Mbc_GR}
$\bcm$ distributions (GR analysis) for data (points) and 
signal (unshaded), $\ddb$ (cross-hatch), continuum (grey), and fake $e^{\pm}$ ($45^{\circ}$ hatch) fit components. (a) $\Dp\ra\eta^{\prime}\enu$ summed over all submodes.  (b--d) $\Dp\ra\eta\enu$ in the indicated $q^{2}$ (GeV$^{2}/c^{4}$) ranges, also
summed over all submodes.}
\end{figure*}

Continuum backgrounds arise largely from $\gamma$ conversions or $\pi^{0}$ Dalitz decays in which one 
$e^{\pm}$ lies below 
identification threshold.  The $e^{\pm}$ candidate is
combined with each track $t$ below the 200 MeV$/c$ threshold yet with $dE/dx$ consistent with an $e^{\mp}$, and each $e^{\pm}t^{\mp}$ pair with every
photon.  Rejecting events with any combination satisfying $m_{e^{+}e^{-}}<100$ MeV$/c^{2}$ or
$|m_{e^{+}e^{-}\gamma}-m_{\pi^{0}}|<50$  MeV$/c^{2}$ almost completely eliminates this background. 

The $\eta^{\prime}\enu$ yields are normalized to the $K^{-}\pi^{+}\pi^{+}$ yield determined using the GR technique,
but with
reversal of the \emissev\ requirement ($\emissev<100$ MeV)
and imposition of a $|\chi_{M}(D)|<3$ requirement on the signal $D$.  Other than the \emissev requirement,
all of the requirements associated with the non-signal $D$ are identical for the semileptonic modes and
the $K^{-}\pi^{+}\pi^{+}$ normalization mode.  As a result, systematic effects associated with the
composition and reconstruction of the second $D$ will largely cancel in the normalization ratio.
 
To find
$q^{2}=(p_{e^{+}}+p_{\nu_{e}})^{2}/c^{2}$, the tagged and GR analyses define the $\nu_{e}$ four momentum $p_\nu$ as
$(E_{\rm miss},  E_{\rm miss}\widehat{\mathbf{p}}_{\rm miss})$ and
$\zeta(|\pmissev|,\pmissev)$, respectively. The $E_{\rm miss}$ calculation in the tagged analysis is 
independent of the tag side. Using the directional information $\widehat{\mathbf{p}}_{\rm miss}$ from
$\mathbf{p}_{\rm miss}$ therefore provides a more uniform calculation of $q^{2}$ across all tag modes.
For the GR analysis, $|\pmissev|$ is determined with better resolution than $\emissev$, leading to
the substitution above for its $q^{2}$ calculation.
The $\etaenu$ data are divided into the $q^{2}$ ranges 
$0 \le q^{2} < 0.5$, $0.5 \le q^{2} < 1.0$, and $q^{2} \ge 1.0$ GeV$^{2}/c^{4}$ to allow study of the form factor.


Efficiency and background determinations utilize a Monte Carlo (MC)
simulation utilizing GEANT~\cite{geant} for the detector
simulation and EvtGen~\cite{Lange:2001uf} for the physics generation.
The analyses utilize
a generic $\ddb$ sample in which both $D$ mesons decay according
to the full model, a
non-$\ddb$ sample that incorporates both continuum $\EE\ra q\bar{q}$ ($q=u, d,~{\rm or}~s$) 
processes and radiative return production of $\psip$, and $\EE\ra\tau^+\tau^-$, as well as
specialized samples for determining signal efficiency with high precision.
The generic $\ddb$ sample is equivalent to 34 times the data statistics.   

Figure~\ref{fig:U} shows the $U$ distributions for the
tagged analysis. We observe
five $\Dp\ra\eta'\enu$ candidates: four events in the
$\etap\ra\pipi\eta$, $\eta\ra\GG$ mode, and one event in the $\etap\ra\pipi\eta$,
$\eta\ra\pipi\piz$ mode.  
Our reconstruction efficiencies, including
subsidiary branching fractions, are $(3.26\pm0.04)\%$ and $(0.86\pm0.02)\%$,
respectively, for these modes.
We expect a total
background for the combined $\etap$ decay modes of $0.043 \pm 0.026$ events.
Our background estimate is based on studies of the $\ddb$ MC sample, of
the generic non-$\ddb$ sample, and of higher statistics MC samples of 
background channels likely to fake  the signal.  The modes 
$\Dp\ra\eta(\pipi\gamma)\enu$, $\Dp\ra\eta(\pipi\piz)\enu$,
and $\Dp\ra\omega(\pipi\piz)\enu$
with correctly identified tags contribute the main $\ddb$ background.
Using a toy simulation that folds Poisson statistics with statistical and 
systematic uncertainties, we find the probability for this background to fluctuate into 5 events to be
$9.7 \times 10^{-9}$, a 5.6 standard deviation (s.d.) significance.  
We find no significant signal for $\phienu$. 

We also search for $\Dp\ra\etap\enu$
with $\etap \ra \rho^0\gamma$ in the tagged analysis. This mode has a large branching
fraction and detection efficiency but also a large background. No
significant signal is observed. A 90 \% C.L. upper limit is set using this decay mode:
${\cal B}(\etapenu) < 3.9 \times 10^{-4}$, which is consistent with the branching fractions
from our observed modes.

The $\eta\enu$ yields are determined from binned likelihood fits to the 
$U$ distributions in each submode.
The signal shape is
described by a modified Crystal Ball function with two power-law
tails~\cite{cb_2tail} that account for initial- and final-state
radiation (FSR) and mismeasured tracks. The signal shape parameters are
fixed to those determined
by fits to signal MC samples. Background function shapes were determined by fitting the $\ddb$ MC sample.
Both normalizations float in the data fits.
The main backgrounds are
misreconstructed semileptonic decays with correctly reconstructed
tags.

\begin{figure*}[tb]
\begin{center}
\includegraphics*[width=10.5cm]{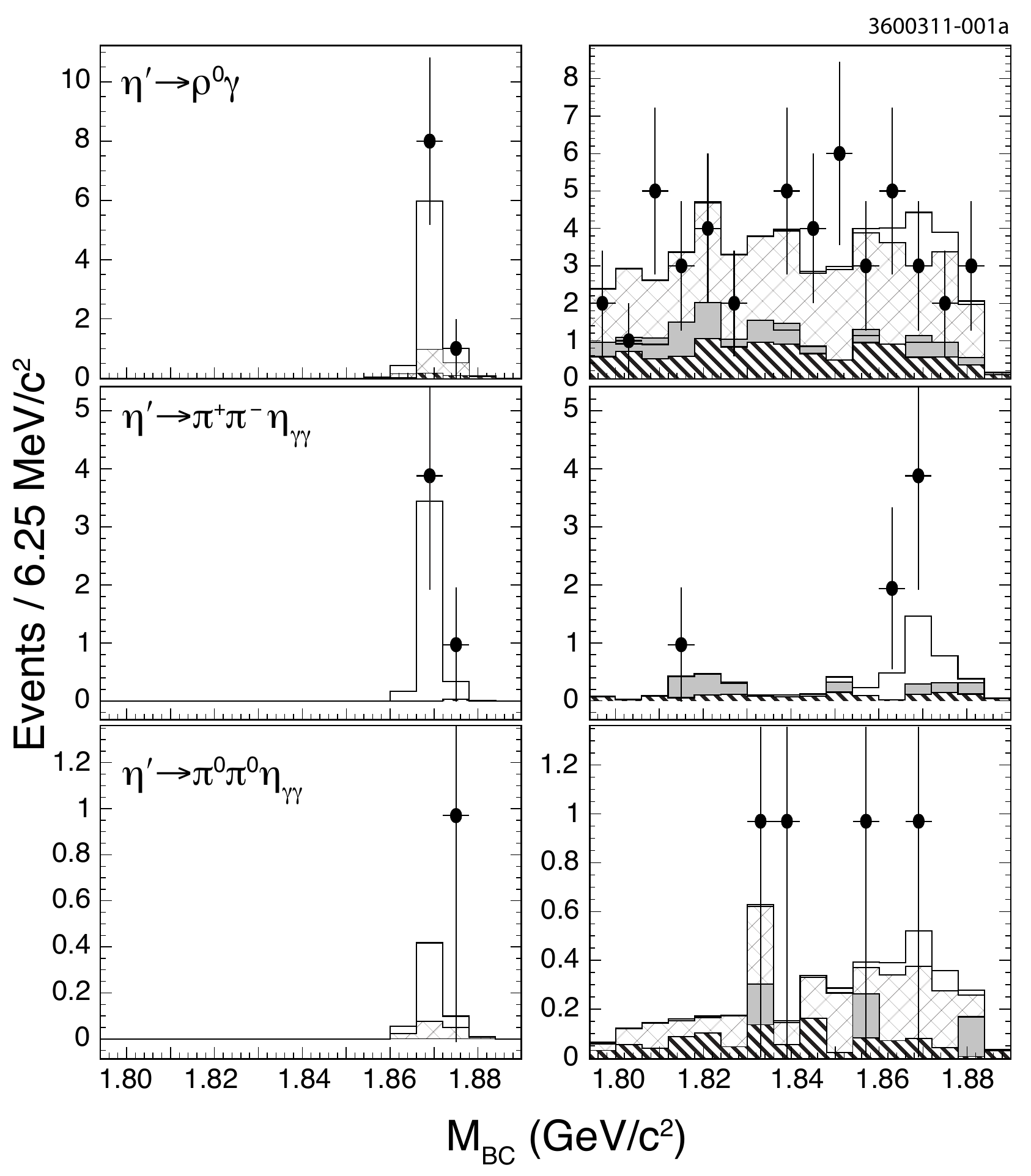}
\caption{\label{fig:etapHQ}
$\bcm$ distributions (GR analysis) for data (points) and 
signal (unshaded), $\ddb$ (cross-hatch), continuum (grey), and fake $e^{\pm}$ ($45^{\circ}$ hatch) fit components for
both the HQ (left) and non-HQ (right) subsamples in the $\eta^{\prime}\to\rho\gamma$ (top), 
$\eta^{\prime}\to\pi^{+}\pi^{-}\eta_{\gamma\gamma}$ (middle), and $\eta^{\prime}\to\pi^{0}\pi^{0}\eta_{\gamma\gamma}$ (bottom)
submodes.}
\end{center}
\end{figure*}

The GR $\bcm$ distributions 
for HQ and non-HQ samples
from all $\etap$ and $\eta$ submodes and $q^{2}$ intervals
and from $K^{-}\pi^{+}\pi^{+}$
are fit simultaneously with reconstructed distributions obtained from the  MC samples for each signal mode, as well as from the generic $\ddb$ and non-$\ddb$ MC samples for background modeling. 
We employ a binned likelihood fit that incorporates the Barlow-Beeston
methodology~\cite{Barlow:1993dm} to accommodate finite MC statistics.
Simultaneous fitting accommodates 
crossfeed among all modes.
The signal and $\ddb$ simulations 
are corrected based on independent data and MC comparisons for the
aspects most critical to the technique: the hadronic $D$ decay model, hadronic showering in the electromagnetic calorimeter, $\pi^{0}$ and $\eta\to\gamma\gamma$
reconstruction efficiencies, $\kl$ energy depositions, and FSR.  To probe the hadronic decay model, we used
the GR reconstruction method with the charged and neutral $D$ hadronic tags $D^{+}\to K^-\pi^+\pi^+$ and $D^0\to K^-\pi^+$, 
respectively, in place of our semileptonic signal modes.  We classified 108 separate decay topologies for the generically-reconstructed $D$ opposite the tag. The observed rates were unfolded and efficiency-corrected, resulting in a decay model that,
when combined with semileptonic measurements, accounts
for $97.2\% \pm 2.0\%$ of all $D$ decays.  To minimize systematic effects in this procedure, the rates were normalized to the unfolded $D^{-}\to K^+\pi^-\pi^-$ and $\bar{D}^0\to K^+\pi^-$ to obtain branching fraction ratios.  These were then
rescaled to world averages~\cite{pdg2008} for these two modes.  As part of this process, we also adjusted daughter 
spectra in the MC to reflect our data.
The efficiency-corrected
$\eta^{\prime}\enu$ and partial $\eta \enu$ yields float in the fit, as does the $\ddb$ background normalization
for each separate submode.  
Figure~\ref{fig:Mbc_GR} shows excellent agreement between data and fit projections.

\begin{figure*}[tb]
\begin{center}
\includegraphics*[width=12cm]{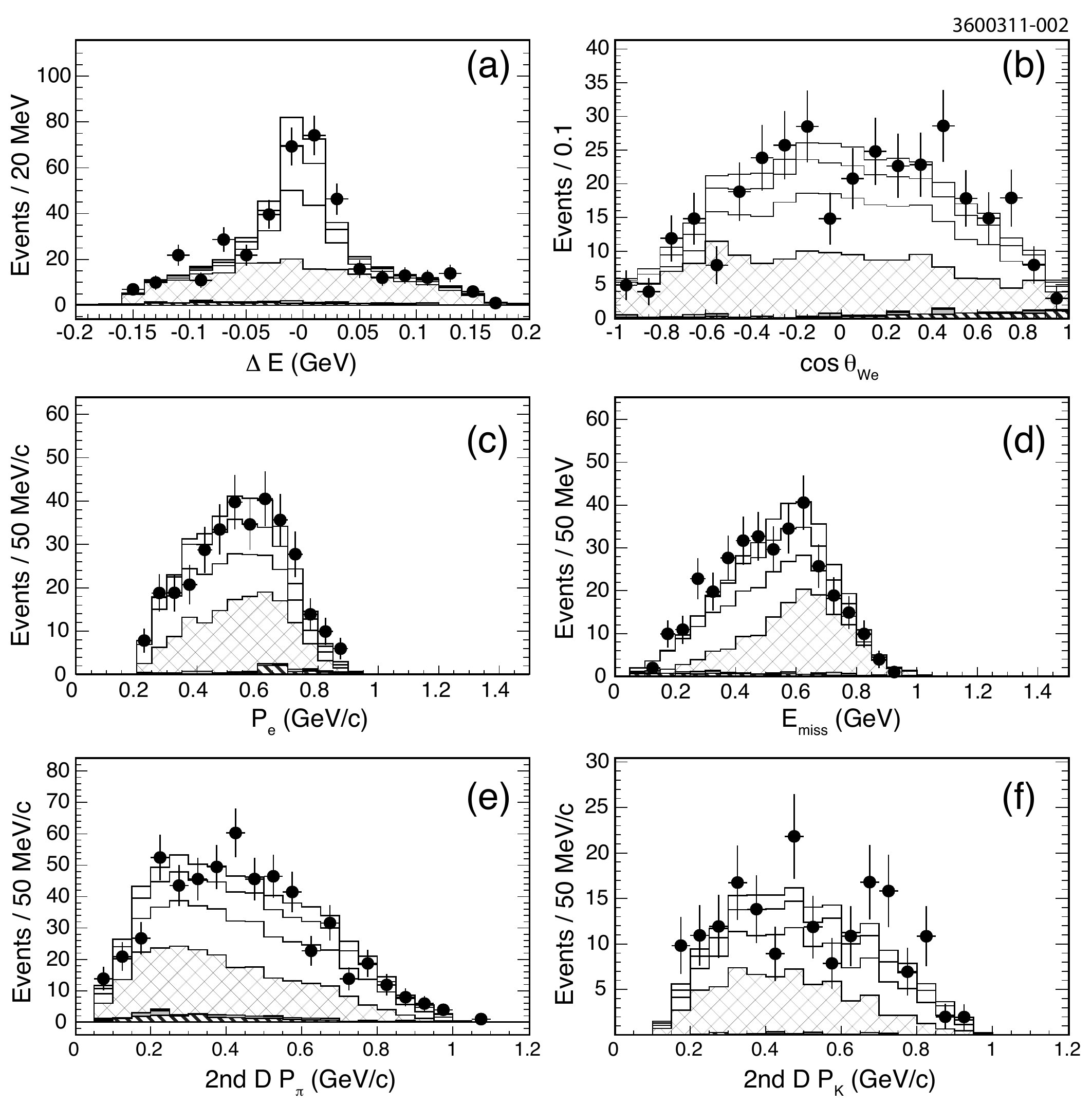}
\caption{\label{fig:etaMisc}
Comparison of data and MC components scaled by nominal GR fit results in the $\etaenu$ mode for
(a) signal side $\Delta E$, (b) $\cos\theta_{We}$, (c) electron momentum spectrum, (d) missing
momentum spectrum, (e) $\pi^{\pm}$ momentum spectrum for the non-signal $D$ side, and 
(f) $\pi^{\pm}$ momentum spectrum for the non-signal $D$ side.  Shown are data (points) and 
signal (unshaded), $\ddb$ (cross-hatch), continuum (grey), and fake $e^{\pm}$ ($45^{\circ}$ hatch) fit components.}
\end{center}
\end{figure*}

\begin{figure*}[tb]
\begin{center}
\includegraphics*[width=12cm]{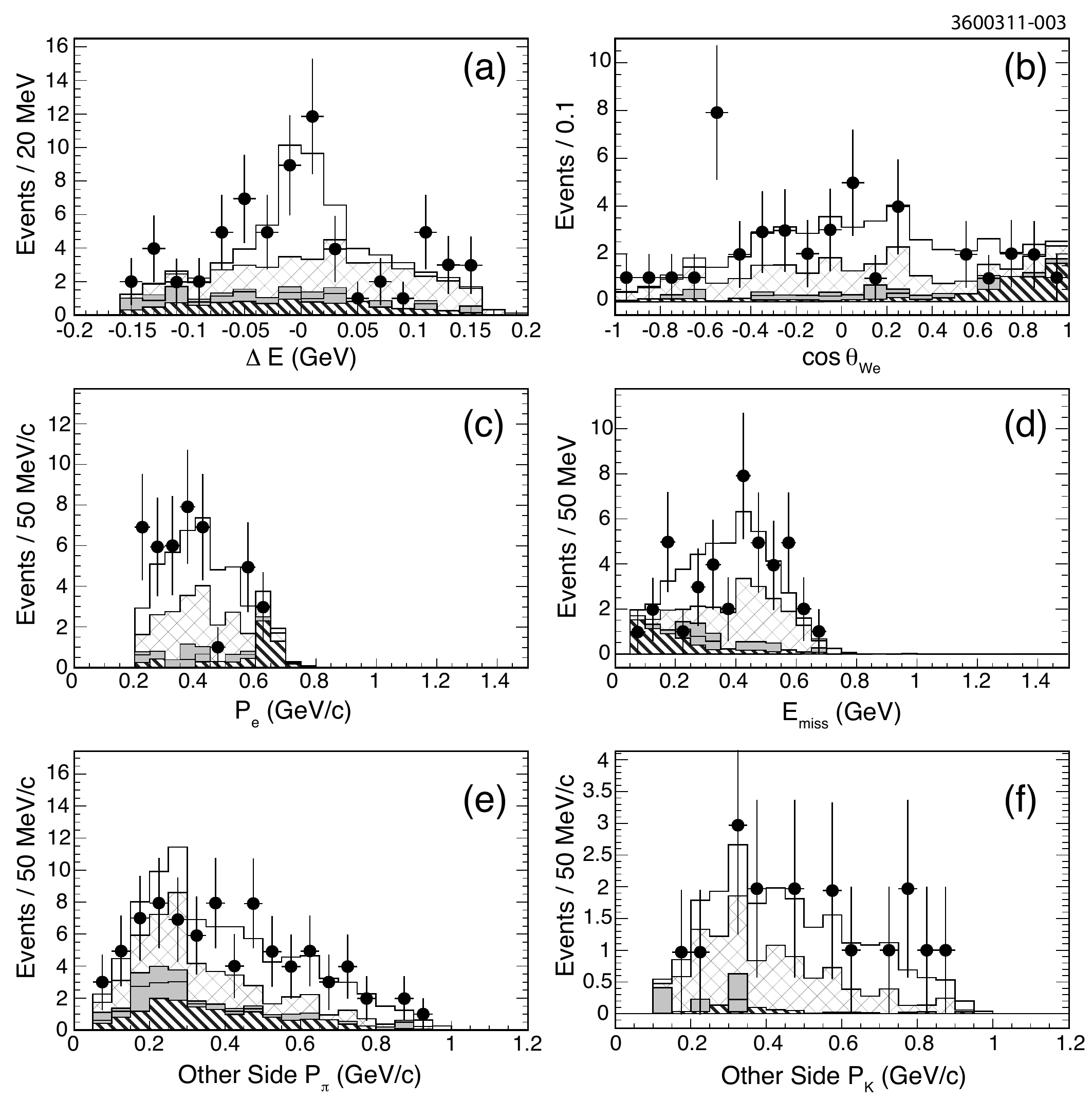}
\caption{\label{fig:etapMisc}
Comparison of data and MC components scaled by nominal GR fit results in the $\etapenu$ mode for
(a) signal side $\Delta E$, (b) $\cos\theta_{We}$, (c) electron momentum spectrum, (d) missing
momentum spectrum, (e) $\pi^{\pm}$ momentum spectrum for the non-signal $D$ side, and 
(f) $\pi^{\pm}$ momentum spectrum for the non-signal $D$ side.  Shown are data (points) and 
signal (unshaded), $\ddb$ (cross-hatch), continuum (grey), and fake $e^{\pm}$ ($45^{\circ}$ hatch) fit components.}
\end{center}
\end{figure*}

The fit likelihood is normalized so that it would correspond to a standard $\chi^{2}$ in the large statistics
limit~\cite{Baker:1983tu}.  We find $-2\ln{\cal L}= 529.7$ for $608-10$ degrees of freedom, providing further evidence of a well-behaved fit.  Fixing the $\eta^{\prime}\enu$ yield at zero
increases the $-2\ln{\cal L}$ by $+33.6$, corresponding to a statistical significance for the observed $\eta^{\prime}\enu$ yield of 5.8 standard deviations.  

The statistical significance given above already incorporates both the background normalization uncertainties and finite MC statistics.  Because the background normalizations float and the background distributions are quite flat, systematic effects must change the $\mbc$ shape to affect the signal yields significantly.  We have used a toy MC simulation to estimate the degradation of the significance from additive systematic effects.  The toy MC model takes the data yields in the $\mbc$ region dominated by signal, integrated over all submodes but subdivided based on the high-quality tagging.  The statistical model includes the independent Poisson fluctuations of the two subsamples, and the background normalization uncertainty that is correlated between the two subsamples.  This toy model, which neglects some information used in the true fit, yields a statistical significance of 5.73 standard deviations, very close to our observed significance.  The additive systematics are dominated by modeling of the $K_{L}^{0}$ energy deposition, of fake charged tracks
and of the momentum spectra of the hadronic decays. When we incorporate the additive systematic uncertainties in the toy MC model, taking into
account the correlations between the two subsamples, we find a reduction in the significance that is
less than 0.05 standard deviations.

The $\eta^{\prime}\enu$ $\bcm$ distributions for the three most influential modes are shown for both 
the HQ and non-HQ samples in Fig.~\ref{fig:etapHQ}.  The data and fit are in excellent agreement across
all these subsamples.  We also examine the signal side $\Delta E$, the electron momentum spectrum, the
missing momentum spectrum and the distribution of $\cos\theta_{We}$ for both $\etaenu$ (Fig.~\ref{fig:etaMisc})
and $\etapenu$ (Fig.~\ref{fig:etapMisc}).  The angle $\theta_{We}$ is the opening angle between the electron
and the virtual $W$ in the $W$ boson's rest frame, and should be distributed as $\sin^{2}\theta_{We}$ for pseudoscalar
to pseudoscalar semileptonic decays such as these.  The MC fit components are scaled according to the nominal fit
results.  The $\Delta E$ range extends outside of the limits imposed for the fit, and none of these distributions
are used in the fit. In both the higher statistics $\etaenu$ and in the $\etapenu$
mode the scaled fit components and data agree very well, but would not without the signal components.  The figures also show
excellent agreement between data and the scaled fit components for the inclusive $\pi^{\pm}$ and $K^{\pm}$ 
momentum spectra from the $D$ reconstructed against the semileptonic candidate.  These comparisons provide
strong support for our observation of $\etapenu$.

\begin{table*}[tb]
\caption{\label{tab:syst} Systematic uncertainties (in percent) for the three $D^+ \rightarrow \eta e^+ \nu_e$ $q^{2}$ intervals
and for the $D^+ \rightarrow \etap e^+ \nu_e$ and $D^+ \rightarrow \phi e^+ \nu_e$ branching fractions.  The $q^{2}$ intervals are quoted in GeV$^{2}/c^{4}$.  The $\eta e^+ \nu_e$ quantities are signed to represent
whether a given uncertainty is correlated or anti-correlated relative to the
corresponding uncertainty for the 0 - 0.5 GeV$^{2}/c^{4}$ $q^{2}$ interval tagged analysis result.}
\begin{tabular}{lddddddddd}
\toprule
                  &\multicolumn{3}{c}{$\eta e^{+} \nu$}             & \multicolumn{3}{c}{$\eta e^{+} \nu$} &\multicolumn{2}{c}{$\etap e^{+}\nu$} & \mcc{$\phi e^{+}\nu$} \\
                  & \multicolumn{3}{c}{tagged}  & \multicolumn{3}{c}{GR} &  \mcc{tagged} & \mcc{GR} & \mcc{tagged}\\
                  & \mcc{0 - 0.5}  & \mcc{0.5 - 1.0}  & \mcc{$\geq 1.0$}
                  & \mcc{0 - 0.5}  & \mcc{0.5 - 1.0}  & \mcc{$\geq 1.0$} \\ 
\colrule
Tracking efficiency                 &  0.40  &  0.42    &  0.43  &  0.11  &  0.12    &  0.10  & 1.06   & 0.01   & 1.30\\
Hadronic identification efficiency  &  0.02  &  0.02    &  0.02  &  1.13  &  1.19    &  1.22  & 0.23   & 0.10   & 0.60\\
$\pi^{0}\to\gamma\gamma$ efficiency &  0.51  &  0.53    &  0.47  &  0.47  &  0.36    &  0.18  & 0.75   & 0.05   &\mcc{--} \\
$\eta\to\gamma\gamma$ efficiency    &  3.03  &  3.07    &  3.27  &  2.23  &  2.42    &  2.87  & 2.93   & 1.29   &\mcc{--} \\
$e^{\pm}$ identification efficiency &  0.66  &  0.54    &  0.43  &  0.66  &  0.54    &  0.43  & 0.70   & 0.70   & 0.80\\
Simulation of FSR                   &  0.13  &  0.06    & -0.15  &  0.50  &  0.76    &  1.27  & 0.30   & 0.50   & 0.30\\
$D^+$ lifetime                      &  0.67  &  0.67    &  0.67  &  0.67  &  0.67    &  0.67  &\mcc{--}&\mcc{--}&\mcc{--} \\
Number of $D$ tags                  &  0.40  &  0.40    &  0.40  &\mcc{--}&\mcc{--}  &\mcc{--}& 0.40   &\mcc{--}& 0.40\\
Tag fakes                           &  0.70  &  0.70    &  0.70  &\mcc{--}&\mcc{--}  &\mcc{--}& 0.70   &\mcc{--}& 0.70\\
U fit Signal Shape                  &  0.37  & -0.50    & -0.52  &\mcc{--}&\mcc{--}  &\mcc{--}&\mcc{--}&\mcc{--}  &\mcc{--} \\
U fit backgrounds                   &  0.64  & -1.09    & -8.25  &\mcc{--}&\mcc{--}  &\mcc{--}& 0.22   &\mcc{--}  &\mcc{--} \\
Simulation of unused tracks         &  0.30  &  0.30    &  0.30  &\mcc{--}&\mcc{--}  &\mcc{--}& 0.30   &\mcc{--}& 0.30\\
Efficiency dependence on $f_{+}(q^{2})$  
                                    &  1.00  & -1.00    & -1.00  &\mcc{--}&\mcc{--}  &\mcc{--}& 1.00   &\mcc{--}& 3.00\\
$q^{2}$ resolution                  &  1.82  & -1.90    & -0.31  &\mcc{--}&\mcc{--}  &\mcc{--}&\mcc{--}&\mcc{--}&\mcc{--} \\
MC statistics                       &  0.88  & -1.10    &  1.55  &\mcc{--}&\mcc{--}  &\mcc{--}& 1.16   &\mcc{--}& 1.60\\
$K_L^0$ showering simulation        &\mcc{--}&\mcc{--}  &\mcc{--}&  0.00  &  0.62    &  1.37  &\mcc{--}&  0.89  &\mcc{--} \\
$K^{\pm}$ identification efficiency &\mcc{--}&\mcc{--}  &\mcc{--}&  0.05  &  0.14    &  0.29  &\mcc{--}& 0.07   &\mcc{--} \\
Fake track simulation               &\mcc{--}&\mcc{--}  &\mcc{--}&  0.02  &  0.04    & -0.08  &\mcc{--}& 0.67   &\mcc{--} \\
Rate of unvetoed hadronic showers   &\mcc{--}&\mcc{--}  &\mcc{--}&  0.58  &  0.05    &  1.80  &\mcc{--}&  0.78  &\mcc{--} \\
Hadronic $D^{+}$ decay model        &\mcc{--}&\mcc{--}  &\mcc{--}&  0.02  &  0.04    & -0.17  &\mcc{--}& 0.04   &\mcc{--} \\
Hadronic $D^{+}$ resonant substructure
                                    &\mcc{--}&\mcc{--}  &\mcc{--}&  0.34  & -0.30    &  0.21  &\mcc{--}&  1.10  &\mcc{--} \\
$K_S^{0}\to \pi^{+}\pi^{-}$ efficiency
                                    &\mcc{--}&\mcc{--}  &\mcc{--}&  0.05  & -0.12    & -0.12  &\mcc{--}&  0.05  &\mcc{--} \\
$K^{\pm}$ as $\pi^{\pm}$ mis-identification
                                    &\mcc{--}&\mcc{--}  &\mcc{--}&  0.02  &  0.01    & -0.00  &\mcc{--}&  0.00  &\mcc{--} \\
${\cal B}(D^+\to K^{-}\pi^{+}\pi^{+})$
                                    &\mcc{--}&\mcc{--}  &\mcc{--}&  2.20  &  2.20    &  2.20  &\mcc{--}& 2.20   &\mcc{--} \\
\botrule
\end{tabular}
\end{table*}

The systematic uncertainties in both analyses are dominated by uncertainties in the
$\eta \ra \GG$ and $\piz\ra\GG$ detection efficiencies, with other
common contributions including track finding efficiency, $e^{\pm}$, $K^{\pm}$
and $\pi^{\pm}$ identification, FSR, and form-factor modeling.  
Efficiency and particle identification uncertainties are determined following
techniques detailed in Ref.~\cite{Dobbs:2007zt}, though modified to reflect
the various selection efficiencies employed by the tagged and GR analyses.
The FSR and form-factor
uncertainty determinations are similar to those in previous semileptonic
analyses~\cite{281etaenu,Nadia}.
Other tagged contributions
include uncertainties in $N_{ \rm tag}$, the no-additional-track requirement,
and the signal $U$ parameterization.  The remaining GR uncertainties arise
in the MC corrections described above.  Many significant uncertainties
({\it e.g.} tracking efficiency and hadronic decay model) 
for the GR analysis
largely cancel in the
$K^{-}\pi^{+}\pi^{+}$ normalization.  
To account for the systematic uncertainty in ${\cal B}(\phienu)$,
we increase the upper limit by one standard deviation.

The systematic uncertainties for both analyses are summarized in Table~\ref{tab:syst}.
The $\eta e^+ \nu_e$ quantities are signed to represent
whether a given uncertainty is correlated or anti-correlated relative to the
corresponding uncertainty for the 0 - 0.5 GeV$^{2}/c^{4}$ $q^{2}$ interval tagged analysis result.
In forming the covariance matrix for the form factor fits for  $\eta e^+ \nu_e$ (see below),
the uncertainties for a given systematic effect are treated either as fully correlated
or anti-correlated.  Treating the uncertainties from each effect as a column vector
$T_{i}$, the covariance matrix $V_{\rm syst}$ is then $V_{\rm syst}=\sum_{i} T_{i}\otimes T_{i}^{\rm T}$.


Table~\ref{br} summarizes all branching fraction and 90\% confidence level (C.L.) upper limit results.
The GR branching fractions ${\cal B}_{\rm GR}$
were obtained from the measured branching ratios
$R_{\rm GR }= {\cal B}(D^{+}\ra\eta^{\prime}\enu)/{\cal B}(D^{+}\ra K^{-}\pi^{+}\pi^{+})$ using 
${\cal B}(D^{+}\ra K^{-}\pi^{+}\pi^{+})=(9.14\pm0.20)\%$~\cite{Dobbs:2007zt}.
The branching fractions measured
using the different $\etap$ and $\eta$ decay modes are consistent in both techniques.  

The tagged and
GR measurements, as well as the partial $\eta\enu$ branching fractions within each measurement, are
statistically and systematically correlated.  To allow proper combination
of the $\eta\enu$ results, we have determined the statistical correlation matrices from an analysis of
event overlap.  Within each analysis, the statistical correlations are obtained from the yield fits.
The statistical correlations, and combined statistical and systematic correlations (see below) are 
summarized in Tables~\ref{tab:statcorr} and~\ref{tab:combinedcorr}, respectively.
The full correlation information is available from EPAPS~\cite{EPAPS} in a machine-readable format
for use in fits by others.

\begin{table}[tb]
\squeezetable
\caption{\label{tab:statcorr} Statistical correlation matrix for the three partial branching fractions
for $\etaenu$ from the two analysis techniques.}
\begin{tabular}{l|ddd|ddd}
\toprule
               &             &  {\rm  tag}&              &              &  {\rm   GR}  &            \\
\colrule
               &          1  &      -0.053   &     -0.002   &       0.43   &          0   &          0 \\
{\rm  tag}     &     -0.053  &           1   &     -0.055   &          0   &       0.39   &          0  \\
               &     -0.002  &      -0.055   &          1   &          0   &          0   &       0.17  \\ \hline
               &       0.43  &           0   &          0   &          1   &    -0.043   &     0.026  \\
{\rm  GR}      &          0  &        0.39   &          0   &    -0.043   &          1   &     -0.022  \\
               &          0  &           0   &       0.17   &     0.026   &     -0.022   &          1  \\
\botrule
\end{tabular}
\end{table}

\begin{table}[tb]
\squeezetable
\caption{\label{tab:combinedcorr} Combined statistical and systematic correlation matrix for the three partial branching fractions
for $\etaenu$ from the two analysis techniques..}
\begin{tabular}{l|ddd|ddd}
\toprule
              &             &{\rm  tag}  &              &              &{\rm  GR} &            \\  
\colrule
              &          1  &      -0.036   &      0.009   &      0.439   &     0.035    &      0.028 \\
{\rm  tag}    &     -0.036  &           1   &     -0.030   &      0.029   &     0.395    &      0.018  \\
              &      0.009  &      -0.030   &          1   &      0.018   &     0.014    &      0.173  \\ \hline
              &      0.439  &       0.029   &      0.018   &          1   &     0.030    &      0.079  \\
{\rm  GR}     &      0.035  &       0.395   &      0.014   &      0.030   &         1    &      0.021  \\
              &      0.028  &       0.018   &      0.173   &      0.079   &     0.021    &          1  \\
\botrule
\end{tabular}
\end{table}

\begingroup
\begin{table}[tb]
\caption{The branching fractions results ${\cal B}_{\rm tag}$ and ${\cal B}_{\rm GR}$ from the tagged
and GR analyses, respectively, and the branching fraction ratios $R_{\rm GR}$
relative to ${\cal B}(D^+ \to K^- \pi^+ \pi^+)$.
The errors are, in order,  the statistical uncertainty and the systematic uncertainty.}
\begin{tabular}{lccc}
\toprule
Mode        & ${\cal B}_{\rm tag}$ [$10^{-4}$] & $R_{\rm GR}$ [\%] & ${\cal B}_{\rm GR}$ [$10^{-4}$] \\
\colrule
$\etap\enu$ & $2.5^{+1.6}_{-1.0}$(0.1)           & $0.237(58)(5)$        & $2.16(53)(7)$ \\
$\phi\,\enu$  & \multicolumn{3}{l}{$<0.9$ @ 90\% confidence level (C.L.)}               \\
$\eta\enu$  & 11.1(1.3)(0.4)                     & 1.28(11)(4)            & $11.7(1.0)(0.4)$ \\
${\eta\enu}_{,0\mhyphen0.5}$   & 6.53(94)(26)              & 0.625(69)(18)          & $5.71(63)(20)$ \\
${\eta\enu}_{,0.5\mhyphen1.0}$ & 3.08(71)(13)              & 0.437(68)(13)          & $3.99(62)(15)$ \\
${\eta\enu}_{,\ge 1.0}$ & 1.77(67)(16)              & 0.223(52)(10)          & $2.03(47)(10)$ \\
\botrule
\end{tabular}
\label{br}
\end{table}
\endgroup

\begin{figure*}[tb]
\includegraphics*[width=10cm]{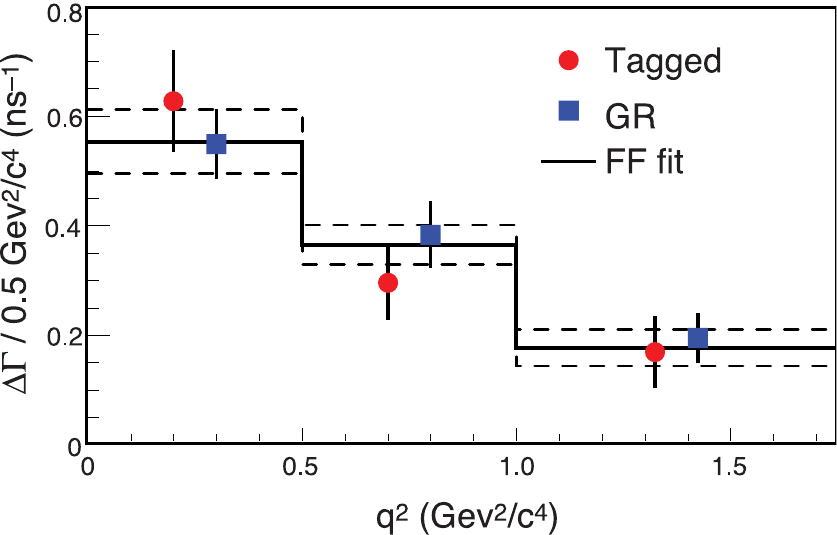}
\caption{
The partial rates from the tagged (circles) and GR (squares) analyses, and
the form factor (FF) fit (histogram).
The dashed lines indicate the total uncertainty on the
fit rates.} \label{fig:comRate}
\end{figure*}

To extract $f_{+}(q^{2})$ for $\etaenu$, we fit the partial rates
obtained from our partial branching fractions
using $\tau_{D^+}=1040(7)\times 10^{-15}{\rm s}$~\cite{pdg2008}.  The fit minimizes
$\chi^{2}=\Delta{\gamma}^{T}{V}^{-1}\Delta{\gamma}$, where
$\Delta{\gamma}={\Delta\Gamma}_{r}-{\Delta\Gamma}_{p}$ is the vector of
differences between the measured ${\Delta\Gamma}_{r}$ and predicted ${\Delta\Gamma}_{p}$
partial widths, 
and ${V}$ is the covariance matrix.
We fit the two analyses both separately and simultaneously, taking into account statistical
correlations from finite $q^{2}$ resolution within an analysis and sample overlap
between analyses.  We fit first with the statistical covariance ${V}={V}_{\rm stat}$, and
then with the combined statistical and systematic covariance ${V}={V}_{\rm stat}+{V}_{\rm syst}$.
The quoted systematic uncertainties are obtained from the quadrature difference of uncertainties
from these two fits.  

We integrate Eq.~(\ref{eq:integ}) over each $q^{2}$ interval to predict ${\Delta\Gamma}_{p}$, parameterizing
the form factor with the standard $z$-expansion parameterization~\cite{becherhill, hillfpcp}
\begin{equation}
f_{+}(q^{2})\equiv \frac{1}{P(q^2)\phi(q^2,t_0)}\sum_{k} a_k z(q^2,t_0)^k.
\end{equation}
We use the standard
form of the outer function $\phi(q^2,t_0)$ and choose $t_{0}$ to minimize the maximum $|z|$ 
over the physical $q^{2}$ range (see Ref.~\cite{becherhill}).  We truncate the series at $k=1$
and allow $f_+(0)\vcd$ and the ratio of linear to constant coefficients, $r_{1}=a_{1}/a_{0}$ to
float in each fit.  This same parameterization was used in our recent measurements of the
$D\to\pi e\nu$ and $D\to K e\nu$ form factors~\cite{818kpienu,Nadia}.

Figure~\ref{fig:comRate} shows the combined fit, and Table~\ref{tab:FF_fit} summarizes the results.
For the combined tagged and GR fit, we find $f_+(0)\vcd = 0.086 \pm 0.006 \pm 0.001$ 
and $r_{1} = -1.83 \pm 2.23\pm 0.28$, with a
correlation of $\rho=0.81$.  The combined fit has a
$\chi^{2}=2.5$ for 4 degrees of freedom.
We obtain the total branching fraction  for the tagged and the combined analyses
by integrating the corresponding fit result.  
Taking $\vcd=0.2256\pm 0.0010$~\cite{pdg2008}, our 
average value for $\vcd f_{+}(0)$ implies
$f_+(0)=0.381\pm0.027 \pm 0.005$. 
Results for other parameterizations of $f_{+}(q^{2})$ are 
discussed in Appendix~\ref{app:otherFits}.

In conclusion, we have made the first observation of the decay mode $D^{+}\ra\etap\enu$
and the first form factor determination for $D^{+}\ra\eta\enu$, as well as improving
its branching fraction measurement.  We also provide the most stringent upper limit on
$D^{+}\ra\phi\enu$ to date.  Our combined branching fraction results are
\begin{eqnarray*}
{\cal B}(D^{+}\ra\eta\enu)  & = & (11.4\pm0.9\pm0.4)\times 10^{-4}, \\
{\cal B}(D^{+}\ra\etap\enu) & = & (2.16\pm0.53\pm0.07)\times 10^{-4}, \\
{\cal B}(D^{+}\ra\phi\enu)  & < & 0.9\times 10^{-4}~~ (90\%~{\rm C.L.}).
\end{eqnarray*}       
These measurements are consistent with our previous results~\cite{281etaenu}, which
they supersede, and with the particle data group's upper
limits~\cite{pdg2008}.  They are also consistent
with \
predictions from both the
ISGW2~\cite{isgw2} and  Fajfer-Kamenic~\cite{fajfer} models.
The upper limit for $\Dp\ra\phi\enu$ is
about twice as restrictive as our previous limit~\cite{281etaenu}.

\begin{table}[t]
\caption{The $D^+\ra\eta\enu $ form factor fit parameters $f_+(0)\vcd$ and $r_{1}$, as well
as their correlation coefficient $\rho$.}
\label{tab:FF_fit}
\begin{center}
\begin{tabular}{lcccc}
\toprule
Analysis    & $f_+(0)\vcd$ & $r_1$             & $\rho$  & $\chi^2/{\rm d.o.f.}$ \\ 
\colrule
Tagged      & 0.094(9)(3)  & 2.17(4.50)(1.12)  & 0.83    & $0.7/(3-2)$  \\
GR          & 0.085(6)(1)  & $-$2.89(2.24)(32) & 0.81    & $0.0/(3-2)$  \\
Combined    & 0.086(6)(1)  & $-$1.83(2.23)(28) & 0.81    & $2.5/(6-2)$ \\
\botrule
\end{tabular}
\end{center}
\end{table}	

\begin{acknowledgments} 
We gratefully acknowledge the effort of the CESR staff 
in providing us with excellent luminosity and running conditions. 
D.~Cronin-Hennessy thanks the A.P.~Sloan Foundation. 
This work was supported by 
the National Science Foundation, 
the U.S. Department of Energy, 
the Natural Sciences and Engineering Research Council of Canada, and 
the U.K. Science and Technology Facilities Council. 
\end{acknowledgments} 

\appendix
\section{Alternate form factor parameterizations}
\label{app:otherFits}

As our primary result for the $D^+ \rightarrow \eta e^+ \nu_e$ form
factor, we utilize the $z$-expansion parameterization \cite{becherhill,hillfpcp} for $f_{+}(q^{2})$.  This
appendix provides fit results for the simple pole and 
Becirevic-Kaidalov~\cite{modpole} (or modified pole) 
parameterizations, which are commonly employed.  The simple pole parameterization takes the form

\begin{equation}
f_+(q^2) = \frac{f_+(0)}{(1-\frac{q^2}{m^2_{\mathrm{pole}}})}, \nonumber
\end{equation}
while the modified pole parameterization takes the form
\begin{equation}
f_+(q^2) = \frac{f_+(0)}{(1-\frac{q^2}{m^2_{D^{*}}})
(1-\alpha\frac{q^2}{m^2_{D^{*}}})}, \nonumber
\end{equation}
where $m_{D^{*}}$ is the $D^{*}$ mass. Either $m_{\mathrm{pole}}$ or
$\alpha$ is fit for along with the form factor zero-intercept.
Table~\ref{tab:oFits} presents the results of a combined fit to the results of
the two analyses using these parameterizations.

\begin{table}[tb]
\caption{\label{tab:oFits} $D^+ \rightarrow \eta e^+ \nu_e$ form factor fit results using the
simple and modified pole parameterizations.  The quantity $\rho$ is the correlation coefficient
between the two fit parameters.  Each fit has $(6-2)$ degrees of freedom.  The shape parameter is
$m_{\mathrm{pole}}$ for the simple pole parameterization and $\alpha$ for the modified pole
parameterization.}
\begin{tabular}{lcc}
\toprule
                & Simple Pole                                & Modified Pole \\
\colrule 
$f^+(0)\vcd$    & $0.086\pm0.005\pm0.001$                    & $0.086\pm0.005\pm0.001$ \\            
shape parameter & $1.87 \pm 0.24\pm0.00$                     & $0.21\pm0.44\pm0.05$ \\
$\rho$          & 0.75                                       & $-0.80$ \\
$\chi^2$        & 2.5                                        & 2.5  \\
\botrule
\end{tabular}
\end{table}

\end{document}